%% file: CDC_v2.tex
\documentclass[letter, 10 pt, conference]{ieeeconf}  

\IEEEoverridecommandlockouts                              
\usepackage{amsmath} 
\usepackage{amssymb}  
\input{math_package}

\input{package_mode}
\usepackage{cite}
\usepackage{psfrag}

\newtheorem{corollary}{Corollary}
\renewcommand{\baselinestretch}{1.125}

\title{\LARGE \bf
A Regularized Saddle-Point Algorithm for Networked Optimization\\ with
Resource Allocation Constraints \\ \textsc{--- Technical Report ---} \\ \normalsize This is an extended version of a paper accepted for CDC 2012 with identical title
}

\author{Andrea Simonetto, Tam\'as Keviczky,  %
\thanks{A.~Simonetto and T.~Keviczky are with the Delft Center for Systems and Control, Delft University of Technology, Mekelweg 2, 2628 CD Delft, The Netherlands, {\tt\footnotesize $\{$a.simonetto, t.keviczky$\}$@tudelft.nl}}%
Mikael Johansson %
\thanks{M.~Johansson is with the ACCESS Linnaeus Centre, School of Electrical Engineering, Royal Institute of Technology (KTH), 100 44 Stockholm, Sweden, {\tt\footnotesize mikaelj@ee.kth.se}}%
}

\begin{document}


\maketitle
\thispagestyle{empty}
\pagestyle{empty}

\begin{abstract}
We propose a regularized saddle-point algorithm for convex networked
optimization problems with resource allocation constraints. Standard
distributed gradient methods suffer from slow convergence and require excessive
communication when applied to problems of this type. Our approach offers
an alternative way to address these problems, and ensures that each
iterative update step satisfies the resource allocation constraints. We
derive step-size conditions under which the distributed algorithm
converges geometrically to the regularized optimal value, and show how
these conditions are affected by the underlying network topology. We
illustrate our method on a robotic network application example where a
group of mobile agents strive to maintain a moving target in the
barycenter of their positions.
\end{abstract}

\input{CDC_main_v2}

\bibliography{../PaperCollection2}
\bibliographystyle{IEEEtran}

\input{appendix_v1}

\end{document}

%% file: math_package.tex
\def\nnorm[#1, #2]{\mbox{$\left. \vline\,\vline\,\displaystyle #1 \,\vline\,\vline \right._{#2}\,$}}

\DeclareMathOperator*{\minimize}{\mathrm{minimize}}
\def\subjto{\mathrm{subject~to}}

\def \xv{x}
\def \r{x_{\mathrm{tot}}}
\def\m{\mu}
\def\La{\mathcal{L}}
\def\alphai{\alpha}
\def\betai{\beta}
\def\kit{\tau}
\def\XV{X}
\def\W{W_{\otimes}}
\def\L{L}
\def\opt{\mathrm{opt}}
\def\qq{t}

%% file: package_mode.tex
\usepackage{amsmath,amsfonts,amssymb}
\usepackage{color}
\usepackage{cite}
\usepackage{graphicx}
\usepackage{algorithm}
\usepackage{algorithmic}
\usepackage{comment}

\def\A[#1, #2]{#1_{#2}}
\def\ti[#1, #2]{#1(#2)}
\def\Ati[#1, #2, #3]{#1_{#2}({#3})}

\newtheorem{theorem}{Theorem}
\newtheorem{assumption}{Assumption}

\newtheorem{lemma}{Lemma}

%% file: CDC_main_v2.tex
\section{Introduction}\label{sec:introduction}

Recent years have witnessed an increased interest in the development of distributed optimization algorithms for diverse applications involving networks of systems, such as multi-robot coordination~\cite{Bullo2008}, distributed estimation~\cite{Maestre2010} , and resource allocation in wireless systems~\cite{Palomar2006}. These algorithms are tailored to accommodate the distributed computation and network constraints inherent in peer-to-peer communication based applications.

In this paper we focus on primal-dual (saddle-point) iterative methods for solving convex optimization problems with resource allocation constraints in addition to convex constraints on local variables and sparse (convex) coupling constraints. Standard (sub)gradient methods \cite{Kiwiel2004, Nedic2009} have gained increased attention in the past years, yet their convergence rate is known to be rather low especially when the cost function is non-strictly convex. A recently proposed method \cite{Devolder2011, Nedic2011} regularizes the initial convex problem and thereby increases the convergence rate of common algorithms delivering a solution arbitrarily close to the one of the original problem. Motivated by this strategy, we also make use of regularization and solve the resulting strictly convex problem via a saddle-point method. Furthermore, we incorporate the resource allocation equality constraints directly into the saddle-point iterations by extending the results of \cite{Xiao2006a} (originally proposed for unconstrained problems). We derive step-size conditions that guarantee convergence of our iterative scheme, and show how these results are linked to the problem characteristics and the graph topology, respectively. In standard dual decomposition approaches \cite{Johansson2008a} one would typically dualize the equality constraints and use consensus mechanisms to distribute the resulting Lagrangian function over the network. This technique results in a slow convergence rate, especially if the network is sparsely connected. Our proposed approach incorporates the resource allocation constraints directly in the saddle-point iteration, and uses regularization to obtain faster convergence. This leads to an inherently distributed method, which converges to the solution of the original regularized problem.

We illustrate our algorithm on a realistic robotic network example where a number of mobile robots strive to keep a moving target in the barycenter of their positions. This scenario is motivated by the recent interest in target tracking and target circumnavigation, e.g.~\cite{Derenick2009,Shames2011}, where distributed algorithms are required to be applicable in real-time.

The paper is organized as follows. Section~\ref{sec:problemformulation} formulates the problem. In Section~\ref{sec:algorithm} we propose the regularized saddle-point algorithm, whose convergence properties are studied in Section~\ref{sec:convergence}. Section~\ref{sec:application} describes the application scenario that is used to illustrate our method. Finally, in Section~\ref{sec:conclusions} we draw our conclusions and plans for future research directions.

\section{Problem Formulation} \label{sec:problemformulation}

We use the standard notation $\mathbf{1}_n$ and $I_n$ to indicate a vector of all ones of dimension $n$ and an identity matrix of dimensions $n\times n$, respectively. The symbol $\otimes$ denotes the Kronecker product, $\langle \cdot, \cdot \rangle$ is the dot product, while $||\cdot||$ is the 2-norm. For a real symmetric matrix $A$, the notation $A \succ 0$ indicates that $A$ is positive definite, $\lambda_i(A)$ denotes its $i$-th smallest eigenvalue whereas the largest one is $\lambda_{\max}(A)$.

We study constrained optimization problems on a network of computing nodes. The network is modeled as a connected graph $\mathcal{G} = (\mathcal{V}, \mathcal{E})$, with vertices (nodes) in the set $\mathcal{V} = \{1, \dots, N\}$ and pairs of nodes as edges in the set $\mathcal{E} \subseteq \mathcal{V}\times\mathcal{V}$. We denote the cardinality of $\mathcal{E}$ as $E$, the set of neighbors of node $i$ as $\mathcal{N}_i = \{j|(i, j) \in \mathcal{E}\}$, while $\L$ is the Laplacian of the graph $\mathcal{G}$, \cite{Godsil2001}.

We consider the following convex constrained optimization problem
\begin{subequations} \label{eq:problem}
\begin{eqnarray} 
\minimize_{\xv_i, \dots, \xv_N} & \sum_{i=1}^N f_i(\xv_i) \label{eq:problemcost} & \\
\subjto & g_{ij}(\xv_i, \xv_j) \leq 0 &\textrm{for all } (i,j) \in \mathcal{E} \label{eq:problemconstr1} \\
        & h_i(\xv_i) \leq 0, &\textrm{for } i = 1, \dots, N \label{eq:problemconstr2} \\
        & \sum_{i = 1}^N \xv_i = \r & \label{eq:problemconstr3}
\end{eqnarray}
\end{subequations}
where each variable $\xv_i \in \mathbb{R}^n$ is associated to the node $i$, each function $f_i: \mathbb{R}^n \to \mathbb{R}$, $g_{ij}: \mathbb{R}^{2n} \to \mathbb{R}$, and $h_{i}: \mathbb{R}^n \to \mathbb{R}$ is a continuously differentiable convex function. The constant vector $\r \in \mathbb{R}^n$ dictates the total amount of each resource that is available in the system and therefore defines the resource allocation constraint\footnote{For simplicity  we consider here the case of $\r \in \mathbb{R}^n$, although our analysis could be extended to handle cases in which $\r \in \mathbb{R}^{\bar{n}}$, with $\bar{n} \leq n$.}. We refer to problem~\eqref{eq:problem} as the primal problem.

Let $\xv \in \mathbb{R}^{nN}$ be the stacked vector $\xv = (\xv_{1}^\top, \dots, \xv_{N}^\top)^{\top}$, while $f(\xv)$ is a separable cost function $f(\xv) = \sum_{i=1}^N f_i(\xv_i)$, and $g(\xv)$ denotes in a compact stacked form all the separable constraint functions described in~\eqref{eq:problemconstr1}-\eqref{eq:problemconstr2},
with codomain (or target set) $\mathbb{R}^m$, $m = E + N$. With this notation we can rewrite problem~\eqref{eq:problem} in the compact form
\begin{subequations}
\begin{eqnarray} \label{eq:vproblem}
\minimize_{\xv} & f(\xv)  \label{eq:vproblemcost} \\
\subjto & \hskip.05cm g(\xv) \leq 0 & \label{eq:vproblemconstr1} \\
        & \hskip1.45cm\left(\mathbf{1}_N^\top \otimes I_n\right) \xv= \r & \label{eq:vproblemconstr2}
\end{eqnarray}
\end{subequations}

We assume that the constraints $g(x) \leq 0$ define a closed and bounded convex set $\bar{\mathbb{\XV}}$. We assume also that the intersection of $\bar{\mathbb{\XV}}$ and the set defined by the equality constraint $\left(\mathbf{1}_N^\top \otimes I_n\right) \xv = \r$ is a closed, bounded, and non-empty convex set, denoted by $\mathbb{\XV}$. Under these assumptions there exists a (possibly non-unique) optimizer of~\eqref{eq:problem}, which we indicate as $\xv^{\opt}$. We denote the primal optimal value by $f^{\opt}$.

Problems of type~\eqref{eq:problem} can be found in various domains including economics~\cite{Arrow1960} and sensor networks~\cite{Palomar2006}. Typically, in these fields the nodes have a coupling resource allocation constraint, for example the total monetary budget or the total available bandwidth, respectively.

We are interested in solving the primal problem~\eqref{eq:problem} via the use of iterative distributed algorithms. 
However, due to the facts that \emph{(i)} the Lagrangian function associated with problem~\eqref{eq:problem} is in general neither strictly convex in the primal variable $\xv$, nor strictly concave in the dual variables,  and \emph{(ii)} the resource allocation constraint couples all the nodes, standard primal-dual iterative methods have typically slow convergence rate and require high communication demand among the nodes. In order to address these issues, we study a regularized version of the saddle-point algorithm (primal-dual iterations) in the next section, which incorporates the resource allocation constraint directly in the update equations.

\section{Regularized Saddle-Point Algorithm}\label{sec:algorithm}

In this section, we present a distributed gradient-based optimization method that employs a fixed regularization in the primal and dual spaces. This regularization serves to approximate the primal problem~\eqref{eq:problem} in a way that can be solved by gradient-based methods with improved convergence properties. Furthermore, we modify the primal iteration to ensure that each iterate satisfies the resource allocation constraint. This allows us to avoid the dualization of the equality constraint, which would need to be distributed among the nodes and lead to increased communication requirements.

Let $\m \in \mathbb{R}^{m}_{+}$ be the dual variable associated to the inequality constraint $g(\xv) \leq 0$, 
and $\nu > 0$, $\epsilon > 0$ be strictly positive scalars. Motivated by \cite{Nedic2011}, we define a regularized Lagrangian-type function associated to the primal problem~\eqref{eq:problem} as
\begin{equation}
\La(\xv, \m) := f(\xv) + \frac{1}{2}\nu ||\xv||^2 + \m^\top g(\xv) - \frac{1}{2}\epsilon ||\m||^2
\label{eq:lagr}
\end{equation}
This Lagrangian-type function is by definition a strictly convex function of the primal variable $\xv$ and a strictly concave function of the dual variable $\m$. 

In order to leverage on strong duality relations, we use the following standard assumption. 
\begin{assumption}
There exists a Slater vector $\bar{\xv} \in \mathbb{X}$ such that $g(\bar{\xv})< 0$.  
\label{ass.slater}
\end{assumption} 

Our aim is to find an (approximate) solution of the primal problem~\eqref{eq:problem}, by solving the regularized saddle-point problem:
\begin{eqnarray}
\max_\m \min_\xv & \La(\xv,\m) \label{eq:reg}\\
\subjto & \left(\mathbf{1}_N^\top \otimes I_n\right) \xv = \r \nonumber
\end{eqnarray}
whose optimal value is denoted by ${f}^*$, and unique optimizer by ${\xv}^*$. Under Assumption~\ref{ass.slater}, the unique optimizer of the regularized problem~\eqref{eq:reg} satisfies the KKT conditions:
\begin{eqnarray*}
 \nabla_\xv \La(\xv^*, \m^*) + \left(\mathbf{1}_N^\top \otimes I_n\right)^\top  p^*  = 0 &\\
 \nabla_\m \La(\xv^*, \m^*)  \leq 0 &\\
 \left(\mathbf{1}_N^\top \otimes I_n\right) \xv - \r = 0 &\\
 \m^*_{q} \nabla_{\m_q} \La(\xv^*, \m^*) = 0 & \textrm{for } q = 1,\dots, M \\
 \m^*_q \geq 0 & \textrm{for } q = 1,\dots, M
\end{eqnarray*}
where $\m^*$ and $p^*$ are the optimal Lagrangian multipliers\footnote{The Lagrangian multiplier $p\in \mathbb{R}^n$ corresponds to the resource allocation equality constraint, which is used to describe the KKT conditions of the regularized problem, but not used in our proposed iterative solution approach.}, while $\nabla_\xv \La$ and $\nabla_\m\La$ indicate the gradients of the regularized Lagrangian-type function $\La(\xv, \m)$ with respect to $\xv$ and $\m$.

It is expected that in general the solutions of the primal problem~\eqref{eq:problem} and the regularized saddle-point problem~\eqref{eq:reg} are different, meaning $||\xv^* - \xv^\opt|| \neq 0$ and $||f^* - f^\opt|| \neq 0$. Furthermore, the solution of the regularized problem~\eqref{eq:reg} does not necessarily satisfy the inequality constraints of the primal problem~\eqref{eq:problem}. However, it is possible to bound the suboptimality and the distance from the primal optimizer, along with the constraint violation by some function of the regularization parameters $\nu$ and $\epsilon$. Thus while we are solving an approximation of the primal problem~\eqref{eq:problem} we have bounds on the distance from the primal optimal solution. Furthermore, in this context the regularization procedure can be seen as a way to speed up the convergence of standard gradient-like methods, which may in fact lead to a closer iterate to the optimum $f^\opt$ of the primal problem within a finite number of iterations even though an approximate regularized problem is being solved. For further details we refer the reader to the original works on regularization and double smoothing techniques~\cite{Devolder2011, Nedic2011}.

The regularized saddle-point problem~\eqref{eq:reg} can be readily solved by centralized iterative methods. However, when a distributed solution is sought, the equality constraint is usually dualized and decomposed among the nodes, see for example the discussions in~\cite{Jadbabaie2009, Zhu2012}.
Typically this procedure causes high communication load and the convergence rate would be affected by the number of nodes. In order to overcome these potential drawbacks we follow a different route and propose an extension to well-known iterative schemes that ensures the feasibility of each iterate with respect to the resource allocation constraint. The main idea can be thought of as ``projecting'' the iterates onto the feasible set of the equality constraint. This extension allows us to design an inherently distributed iterative scheme that still solves the original regularized problem~\eqref{eq:reg}.

Let $\mathcal{P}_{\mathbb{R}_+}$ indicate the projection over the positive orthant, and let $\alphai >0$ and $\betai > 0$ be fixed strictly positive scalars (step sizes). We consider the following saddle-point iterations:
\begin{eqnarray}
\xv^{(\kit + 1)} &=& \xv^{(\kit)} - \alphai \betai  (W \otimes I_n) \nabla_x\La(\xv^{(\kit)}, \m^{(\kit)}) \label{eq:itx}\\
\m^{(\kit + 1)} &=& \mathcal{P}_{\mathbb{R}_+}\left[\m^{(\kit)} + \alphai {\nabla_\m\La}(\xv^{(\kit)}, \m^{(\kit)})\right] \label{eq:itmu}
\end{eqnarray}
with any matrix $W \in \mathbb{R}^{N \times N}$ such that
\begin{enumerate}
\item[\emph{(a)}] the vectors $\mathbf{1}_N$ and $\mathbf{1}_N^\top$ are left and right eigenvectors of $W$ associated to the zero eigenvalue, respectively:
$$
\mathbf{1}_N^\top W = \mathbf{0},\qquad W \mathbf{1}_N = \mathbf{0}
$$
\item[\emph{(b)}] the zero eigenvalue is unique, i.e.,
$$
W + W^\top + (1/N) \mathbf{1}_N\mathbf{1}_N^\top \succ 0
$$
\item[\emph{(c)}] the matrix $W$ has the same sparsity pattern as the Laplacian matrix $\L$ of the graph $\mathcal{G}$.

\end{enumerate}
It is easy to see that if the properties \emph{(a)-(c)} hold,
then the iterations~\eqref{eq:itx}-\eqref{eq:itmu} can be computed locally with only the information of the neighboring nodes. In this sense, the iterations~\eqref{eq:itx}-\eqref{eq:itmu} are inherently distributed.

We claim that there exist conditions on the step-sizes $\alphai$ and $\betai$ such that the iterations~\eqref{eq:itx}-\eqref{eq:itmu} converge to the unique optimal solution of the regularized problem~\eqref{eq:reg}. In particular, we expect that the step size $\alphai$ is linked to the characteristics of the functions $f$ and $g$ (as in standard gradient-like methods), while $\betai$ is linked to $W$, i.e., the network topology. These relationships will be shown using the following lemma, which establishes three important properties of the iterations~\eqref{eq:itx}-\eqref{eq:itmu}.
\begin{lemma} If the matrix $W$ satisfies the property \emph{(a)} and for the first iterate $\xv^{(0)}$ the resource allocation constraint holds, i.e., $(\mathbf{1}_N^\top \otimes I_n) \xv^{(0)} = \r$, then
\begin{enumerate}
\item[\emph{(i)}] for any $\kit$, the iterate $\xv^{(\kit)}$ satisfies the resource allocation constraint, i.e., $(\mathbf{1}_N^\top \otimes I_n) \xv^{(\kit)} = \r$;
\item[\emph{(ii)}] the optimal couple $(\xv^*, \m^*)$ of~\eqref{eq:reg} is a fixed point of the iterations~\eqref{eq:itx}-\eqref{eq:itmu};
\item[\emph{(iii)}] for any $\kit$, the equality $\xv^{(\kit+1)} = \xv^{(\kit)}$ holds if and only if either $\nabla_x\La(\xv^{(\kit)}, \m^{(\kit)}) + (\mathbf{1}_N \otimes I_n) p = 0$, for some $p\in \mathbb{R}^n$, or $\nabla_x\La(\xv^{(\kit)}, \m^{(\kit)}) = \mathbf{0}$.
\end{enumerate}
\label{lemma.xiao}
\end{lemma}
\emph{Proof}.
The first claim follows by induction based on~\cite{Xiao2006a}. Suppose that $\xv^{(\kit)}$ satisfies the resource allocation constraint. Then for $\xv^{(\kit+1)}$
\begin{multline*}
\left(\mathbf{1}_N^\top \otimes I_n\right)\xv^{(\kit + 1)} =\\ \left(\mathbf{1}_N^\top \otimes I_n\right)\left(\xv^{(\kit)} - \alphai \betai  (W \otimes I_n) \nabla_x\La(\xv^{(\kit)}, \m^{(\kit)})\right)
\end{multline*}
and using the fact that $\left(\mathbf{1}_N^\top \otimes I_n \right)\left(W \otimes I_n\right) = \mathbf{1}_N^\top W \otimes I_n = \mathbf{0}$ (property \emph{(a)} of $W$) the claim follows.

The second claim follows by direct calculations. Consider the optimal pair $(\xv^*, \m^*)$ of~\eqref{eq:reg}, then using the KKT conditions we obtain
\begin{multline*}
\xv^{(\kit+1)} = \xv^* - \alpha \beta (W \otimes I_n) \nabla_\xv \mathcal{L}(\xv^*, \m^*) =\\ \xv^* + \alpha \beta \left(W \otimes I_n\right) \left(\mathbf{1}_N \otimes I_n\right)  p^*.
\end{multline*}
Since $\left(W \otimes I_n \right)\left(\mathbf{1}_N \otimes I_n\right) =  W \mathbf{1}_N \otimes I_n = \mathbf{0}$, it follows that $\xv^{(\kit+1)} = \xv^*$ and therefore $\xv^*$ is a fixed point.

The third claim follows from property \emph{(b)} of $W$, i.e., the uniqueness of the zero eigenvalue. The equality $\xv^{(\kit + 1)} = \xv^{(\kit)}$ holds if and only if $\alphai \betai  (W \otimes I_n) \nabla_x\La(\xv^{(\kit)}, \m^{(\kit)}) = \mathbf{0}$. This last equality is true either if $\nabla_x\La(\xv^{(\kit)}, \m^{(\kit)}) = \mathbf{0}$ or if the vector $\nabla_x\La(\xv^{(\kit)}, \m^{(\kit)})$ is an eigenvector of $W$ with associated zero eigenvalue. Therefore, using property \emph{(b)} of $W$ leads to $\nabla_x\La(\xv^{(\kit)}, \m^{(\kit)}) = (\mathbf{1}_N \otimes I_n) p'$, with $p' \in \mathbb{R}^n$. Choosing $p' = -p$ proves the claim.
\hfill $\Box$

Lemma~\ref{lemma.xiao} shows that the matrix $W$ keeps the iterates feasible with respect to the resource allocation constraint and does not introduce undesired fixed points.

The next section investigates the conditions on $\alphai$ and $\betai$ under which the primal-dual iterates $\xv^{(\kit)}$ and $\m^{(\kit)}$ converge to the optimizer $(\xv^*, \m^*)$ of~\eqref{eq:reg}, and the bounds on how far this solution is from the primal solution $\xv^\opt$ in terms of suboptimality $||f^* - f^\opt||$ and constraint violation.

\section{Convergence Properties}\label{sec:convergence}

Let $z$ be the stacked vector $z = (\xv^\top, \m^\top)^\top$, and define the mapping $\Phi(z) = (\nabla_\xv \La(z)^\top, -\nabla_\m\La (z)^\top)^\top$. We use the short-hand notation $\W = W \otimes I_n$. Moreover, let $\mathcal{P}$ be a generic projection operator, whose codomain will be clear by the context. The iterations~\eqref{eq:itx}-\eqref{eq:itmu} can be compactly written as:
\begin{equation}
z^{(\kit + 1)} = \mathcal{P}\left[z^{(\kit)} -\alpha \left[ \begin{array}{cc} \beta \W & \\ & I_M\end{array}\right]\Phi(z^{(\kit)})\right] =: T(z^{(\kit)})
\end{equation}

The scope of this section is to identify the assumptions on $\La(\xv, \m)$ and the conditions on $\alphai$ and $\betai$ that let the mapping $T: \mathbb{R}^{Nn}\times \mathbb{R}^{m}_{+} \to \mathbb{R}^{Nn}\times \mathbb{R}^{m}_{+}$ be a contraction mapping. This guarantees geometric convergence of the iterations~\eqref{eq:itx}-\eqref{eq:itmu} to the optimal point of~\eqref{eq:reg}.

First of all we characterize the properties of the mapping $\Phi(z)$ under the following standard assumptions.

\begin{assumption}\label{ass.1}
The iterates $\xv^{(\kit)}$ and $\m^{(\kit)}$ are contained in some closed, convex, and bounded sets for each iteration $\kit$. In other words, $\xv^{(\kit)} \in \hat{\mathbb{\XV}}$ and $\m^{(\kit)} \in \hat{\mathbb{M}}$.
\end{assumption}
We note that the assumption of $\m^{(\kit)} \in \hat{\mathbb{M}}$ is satisfied under Assumption~\ref{ass.slater} (see~\cite{Nedic2011} for details).

On the other hand, the assumption on $\xv^{(\kit)}$ can be difficult to justify. Although if the sequence $\{\xv^{(\kit)}\}$ converges to $\xv^*$ we know that $\xv^{(\kit)}$ is asymptotically bounded, there is no guarantee a priori that $\xv^{(\kit)}$ is bounded in each iteration. Furthermore, we cannot simply project the $\xv$ iterates in \eqref{eq:itx} on some closed, convex, and bounded set, since this would destroy the properties of the information exchange matrix $W$. In the Appendix 
we show a simple way to locally modify the function $\L(\xv, \m)$ in order to \emph{enforce} that $\xv^{(\kit)} \in \hat{\mathbb{\XV}}$, and therefore ensure the boundedness of the iterates $\xv^{(\kit)}$. Since this local modification does not change our converge analysis, we assume now that Assumption~\ref{ass.1} is satisfied and we refer to the Appendix for the technical details.

We make the following mild and technical assumptions:
\begin{assumption}\label{ass.2}
The gradients of $f(\xv)$ and each $g_{q}(\xv)$ are Lipschitz continuous with constants $F$ and $G_{q}$, respectively:
\begin{equation*}
\begin{array}{rl}
||\nabla_\xv f(a) - \nabla_\xv f(b)|| \leq  F ||a - b||, &\textrm{ for } a, b \in\hat{\mathbb{\XV}}\\
||\nabla_\xv g_{q}(a) - \nabla_\xv g_{q}(b)|| \leq G_q ||a - b||, &
\textrm{ for } a, b \in\hat{\mathbb{\XV}},\\ & q = 1,\dots,m
\end{array}
\end{equation*}
\end{assumption}
\begin{assumption}\label{ass.3}
The constraint gradient and Lagrangian dual variable are bounded as
\begin{equation*}
||\nabla_\xv g(\xv)|| < M_d, \textrm{ and } ||\m||\leq M_{\mu}.
\end{equation*}
\end{assumption}
We note that Assumptions~\ref{ass.2} and \ref{ass.3} are commonly required in the analysis of gradient descent methods. Furthermore, Assumption~\ref{ass.3} is generally satisfied under Assumption~\ref{ass.1}.

Assumptions~\ref{ass.slater}, \ref{ass.1}, \ref{ass.2}, and \ref{ass.3} are important to guarantee that the mapping $\Phi(z)$ has certain regularity properties. In fact, under these assumptions, by Lemma~3.4 of~\cite{Nedic2011}, the mapping $\Phi(z)$ is strongly monotone with constant $\varphi = \min(\nu, \epsilon)$ and Lipschitz with constant $F_\Phi$. In other words,
\begin{equation}\label{eq:mon}
\left\langle \Phi(a) - \Phi(b), a - b  \right\rangle \geq \varphi ||a - b||^2, \, \textrm{ for } a, b \in \hat{\mathbb{\XV}}\times\hat{\mathbb{M}}
\end{equation}
\begin{equation}\label{eq:Lip}
||\Phi(a) - \Phi(b)|| \leq F_{\Phi} ||a - b||, \, \textrm{ for } a, b \in \hat{\mathbb{\XV}}\times\hat{\mathbb{M}}
\end{equation}
Properties~\eqref{eq:mon} and \eqref{eq:Lip} will be important for convergence.

\subsection{Symmetric Case}

In this subsection we will assume that the matrix $W$ is symmetric, i.e., $W^\top = W$. This will allow us to derive closed-form conditions for the step-sizes $\alphai$ and $\betai$. Define
\begin{equation}
C := \max(\betai \lambda_{\max}(W), 1)
\label{eq:C}
\end{equation}
\begin{equation}
\kappa := \varphi - {F_{\Phi}}  \left(\betai \lambda_{\max}(W)  - 1\right)
\label{eq:kappa}
\end{equation}

\begin{theorem}\label{theo:conv}
Under the Assumptions~\ref{ass.slater}, \ref{ass.1}, \ref{ass.2}, and \ref{ass.3} and for symmetric $W$, the conditions
\begin{equation}
\betai \lambda_{\max}(W) < 1 + \frac{\varphi}{F_{\Phi}}, \quad \textrm{ and } \quad \alphai < \frac{2 \kappa }{C^2 F_{\Phi}^2}
\label{eq:conditions}
\end{equation}
ensure geometrical convergence of the iterations~\eqref{eq:itx}-\eqref{eq:itmu} to the unique optimizer $(\xv^*, \m^*)$ of the regularized problem~\eqref{eq:reg}. Furthermore, the convergence rate $r$ is
\begin{equation}
r = 1 - 2 \alpha \kappa + \alpha^2 C^2 F^2_{\Phi}
\label{eq.convergencerate}
\end{equation}
\end{theorem}
\emph{Proof.} The distance of the primal iterate $\xv^{(\kit + 1)}$ to a primal optimizer $\xv^*$ can be written as
\begin{multline}
||\xv^{(\kit + 1)} - \xv^*||^2 = ||\xv^{(\kit)} - \xv^*||^2 -\\ 2\alphai \left \langle \betai \W \left(\nabla_\xv \La(\xv^{(\kit)}, \m^{(\kit)}) - \nabla_\xv\La(\xv^*, \m^*)\right), \xv^{(\kit)} - \xv^* \right \rangle  +\\  \alphai^2 \left\|\betai \W \left(\nabla_\xv \La(\xv^{(\kit)}, \m^{(\kit)})- \nabla_\xv \La(\xv^*, \m^*)\right)\right\|^2
\label{eq:relx0}
\end{multline}
where we made use of the fact that $\xv^*$ is a fixed point of the iteration~\eqref{eq:itx}. Using the relation
\begin{multline*}
\left\|\betai \W \left(\nabla_\xv \La(\xv^{(\kit)}, \m^{(\kit)})- \nabla_\xv\La(\xv^*, \m^*)\right)\right\|^2 \leq \\ \betai^2\lambda_{\max}^2(W)\left\|\nabla_\xv \La(\xv^{(\kit)}, \m^{(\kit)})- \nabla_\xv \La(\xv^*, \m^*)\right\|^2
\end{multline*}
equation~\eqref{eq:relx0} becomes
\begin{multline}
||\xv^{(\kit + 1)} - \xv^*||^2 \leq ||\xv^{(\kit)} - \xv^*||^2 - \\ 2\alphai \left \langle \betai \W \left(\nabla_\xv \La(\xv^{(\kit)}, \m^{(\kit)})- \nabla_\xv\La(\xv^*, \m^*)\right), \xv^{(\kit)} - \xv^* \right \rangle  +\\  \alphai^2 \betai^2 \lambda_{\max}^2(W) \left\|\nabla_\xv \La(\xv^{(\kit)}, \m^{(\kit)})- \nabla_\xv\La(\xv^*, \m^*)\right\|^2
\label{eq:relx}
\end{multline}
In a similar fashion, and using the non-expansive property of the projection, we can write the distance of the Lagrangian multiplier $\m^{(\kit + 1)}$ to its optimal value $\m^*$ as:
\begin{multline}
||\m^{(\kit + 1)} - \m^*||^2 \leq ||\m^{(\kit)} - \m^*||^2 + \\ 2\alphai \left \langle \nabla_\m \La(\xv^{(\kit)}, \m^{(\kit)})- \nabla_\m \La(\xv^*, \m^*), \m^{(\kit)} - \m^* \right\rangle  +\\ \alpha^2 \left\|\nabla_\m \La(\xv^{(\kit)}, \m^{(\kit)})- \nabla_\m \La(\xv^*, \m^*)\right\|^2
\label{eq:relmu}
\end{multline}
Summing up the relations~\eqref{eq:relx}-\eqref{eq:relmu} we obtain
\begin{multline}
||z^{(\kit + 1)} - z^*||^2 \leq ||z^{(\kit)} - z^*||^2 - \\ 2\alpha \left\langle \left[ \begin{array}{cc}\beta \W & \\ & I_M\end{array}\right]\left(\Phi(z^{(\kit)})- \Phi(z^*)\right), z^{(\kit)} - z^* \right\rangle  +\\ \alpha^2 C^2 \left\|\Phi(z^{(\kit)})- \Phi(z^*))\right\|^2
\label{eq:zit}
\end{multline}
where $C$ is defined as in~\eqref{eq:C}. The term
\begin{equation*}
-\left\langle \left[ \begin{array}{cc}\beta \W & \\ & I_M\end{array}\right]\left(\Phi(z^{(\kit)})- \Phi(z^*)\right), z^{(\kit)} - z^* \right\rangle
\end{equation*}
can be expanded as
\begin{multline*}
-\left\langle \left[ \begin{array}{cc}\beta \W & \\ & I_M\end{array}\right]\left(\Phi(z^{(\kit)})- \Phi(z^*)\right), z^{(\kit)} - z^* \right\rangle = \\ \underbrace{-\left\langle \left[ \begin{array}{cc}I_{Nn} & \\ & I_M\end{array}\right]\left(\Phi(z^{(\kit)})- \Phi(z^*)\right), z^{(\kit)} - z^* \right\rangle}_{\textrm{(a)}} +\\
\underbrace{-\left\langle \left[ \begin{array}{cc}\betai \W - I_{Nn} & \\ & 0\end{array}\right]\left(\Phi(z^{(\kit)})- \Phi(z^*)\right), z^{(\kit)} - z^*\right\rangle}_{\textrm{(b)}}
\end{multline*}
We can bound the term~(a) based on the strong monotonicity of $\Phi(z)$ in~\eqref{eq:mon}, while the term~(b) can be bounded as
\begin{multline*}
\hskip-7pt-\left\langle \left[ \begin{array}{cc}\betai \W - I_{Nn} & \\ & 0\end{array}\right]\left(\Phi(z^{(\kit)})- \Phi(z^*)\right), z^{(\kit)} - z^*\right\rangle
  \\ \hskip-2.5pt
\leq \left\|\left\langle \left[ \begin{array}{cc}\betai \W - I_{Nn} & \\ & 0\end{array}\right]\left(\Phi(z^{(\kit)})- \Phi(z^*)\right), z^{(\kit)} - z^* \right\rangle \right\|   \\
\leq \left\|\left[ \begin{array}{cc}\betai \W - I_{Nn} & \\ & 0\end{array}\right]\right\| \left\|\Phi(z^{(\kit)})- \Phi(z^*)\right\|\left\| z^{(\kit)} - z^* \right\| \\ \leq
\left(\betai \lambda_{\max}(W) - 1\right) {F_{\Phi}} ||z^{(\kit)} - z^*||^2
\end{multline*}
where we used the Lipschitz continuity of $\Phi(z)$ in~\eqref{eq:Lip}. The relation~\eqref{eq:zit} then becomes
\begin{equation}
||z^{(\kit + 1)} - z^*||^2 \leq \left(1 - 2 \alpha \kappa + \alpha^2 C^2 F^2_{\Phi}\right)||z^{(\kit)} - z^*||^2 \label{eq:zit2}
\end{equation}
with $\kappa$ defined as in~\eqref{eq:kappa}. Therefore the first convergence condition is $1 - 2 \alpha \kappa + \alpha^2 C^2 F^2_{\Phi} < 1$, while, since it is required that $\alphai > 0$, the second condition must be $\kappa > 0$. From these two conditions the relations~\eqref{eq:conditions} follow. Furthermore the convergence rate expression in~\eqref{eq.convergencerate} can be established based on~\eqref{eq:zit2}. \hfill $\Box$

\begin{corollary}\label{cor:small}
The convergence conditions~\eqref{eq:conditions} on the step sizes can be upper-bounded by
\begin{equation}
  \beta < \frac{1}{\lambda_{\max}(W)}, \quad \textrm{ and }\quad \alpha < 2 \frac{\varphi}{F^2_{\Phi}}
  \label{eq:corollary}
\end{equation}
\end{corollary}
\emph{Proof.} The proof follows directly from $\varphi/F_{\Phi} > 0$.

The type of conditions in Corollary~\ref{cor:small} are typical in (sub)gradient methods and are often referred to as ``small enough'' step size conditions~\cite{Bertsekas1999}. We may notice that $\alpha$ is bounded by quantities related to the characteristics of the problem functions, while $\beta$ is related to the structure of the information exchange graph. We also note that $\alphai$ has to be determined a priori based on the knowledge of the problem function properties, while $\betai$ can be computed in a distributed way by the nodes, since there are distributed algorithms to upper-bound $\lambda_{\max}(W)$, e.g.~\cite{Li2001}.

The following technical lemma characterizes the ``quality'' of the regularized optimal solution $\xv^*$ with respect to the original primal problem~\eqref{eq:problem}: it provides bounds on the amount of constraint violation of $g(\xv^*)$ and the suboptimality $||f^{\mathrm{opt}} - f^*||$. In order to compactly characterize these bounds, we define the constraint set of the regularized problem~\eqref{eq:reg} as $\mathbb{\XV}_{\nu,\epsilon}$, which implies that $\xv^* \in \mathbb{\XV}_{\nu,\epsilon}$. The set $\mathbb{\XV}_{\nu,\epsilon}$ is closed, bounded, and convex, and in general different from the original primal constraint set $\mathbb{\XV}$, being however $\mathbb{X} \subseteq \mathbb{X}_{\nu, \epsilon}$.

\begin{lemma} Under the Assumptions~\ref{ass.slater}, \ref{ass.1}, \ref{ass.2}, and \ref{ass.3} the maximum constraint violation is bounded by
\begin{equation}\label{eq:constviol}
\max\{0, g_i(\xv^*)\} \leq M_{d_i} M_{\mu} \sqrt{\frac{\epsilon}{2 \nu}}
\end{equation}
where $M_{d_i} = \max_{x \in \hat{\mathbb{\XV}}}||\nabla_\xv g_i(\xv)||$ for each $i$ and $M_{\mu} = \max_{\m \in \hat{\mathbb{M}}} ||\m||$, while the difference between the optimal value of the regularized problem~\eqref{eq:reg} and the optimal value of the original one~\eqref{eq:problem} can be bounded by
\begin{equation}
||f^* - f^\opt|| \leq M_f M_{\mu} \sqrt{\frac{\epsilon}{2 \nu}} + \frac{\nu}{2} D^2
\label{eq:bound2}
\end{equation}
where $M_f = \max_{\xv \in {\mathbb{\XV}}_{\nu, \epsilon}} ||\nabla_\xv f(\xv)||$, $D = \max_{\xv \in {\mathbb{\XV}}_{\nu, \epsilon}} ||\xv||$
\end{lemma}

\emph{Proof.}  The proof is a modified version of Lemma 3.3 in~\cite{Nedic2011}. In particular, the bound~\eqref{eq:constviol} follows directly from Lemma 3.3 in~\cite{Nedic2011}, while the bound~\eqref{eq:bound2} requires the modification that we consider for $M_f$ and $D$ the maximum of $x$ over $\mathbb{\XV}_{\nu, \epsilon}$ instead of $\hat{\mathbb{\XV}}$ (which could lead to a too conservative result in our case). 

The proof of the bound~\eqref{eq:bound2} starts from bounding $||f^* - f^{\opt} ||$ by 
\begin{equation}
||f^* - f^{\opt} ||\leq ||f^* - f^*_{\epsilon=0}|| + f^*_{\epsilon=0} - f^\opt
\label{eq.tech1}
\end{equation} 
where $f^*_{\epsilon=0}$ is the optimal cost for the regularization problem with regularization parameter $\epsilon = 0$, and $f^*_{\epsilon=0} - f^\opt \geq 0$. By convexity of $f$, we have
\begin{equation}
f^* - f^*_{\epsilon=0}\leq \nabla_x f(x^*)^\top (x^* - x^*_{\epsilon=0}) \label{eq.tech2}
\end{equation} 
with $x^*_{\epsilon=0}$ the unique optimizer of the regularization problem with regularization parameter $\epsilon = 0$. Since $\xv^*,  x^*_{\epsilon=0} \in \mathbb{X}_{\nu, \epsilon}$ and $\mathbb{X}_{\nu, \epsilon}$ is compact, by the continuity of the gradient $||\nabla_x f(x)||$, the gradient norm is bounded and we can write~\eqref{eq.tech2} as  
\begin{equation}
||f^{*} - f^{*}_{\epsilon=0}||\leq \underbrace{\max_{x \in \mathbb{X}_{\nu, \epsilon}}||\nabla_x f(x)||}_{M_f} ||x^* - x^*_{\epsilon=0}|| \label{eq.tech3}
\end{equation}
However, By Proposition 3.1 of \cite{Nedic2011}, we can bound $||x^* - x^*_{\epsilon=0}||$ by $M_\mu \sqrt{\epsilon/2\nu}$ and therefore~\eqref{eq.tech3} can be written as
\begin{equation}
||f^* - f^*_{\epsilon=0}||\leq M_f M_\mu \sqrt{\frac{\epsilon}{2 \nu}}\label{eq.tech4}
\end{equation}
Using the estimate $f^*_{\epsilon=0} - f^\opt \leq \nu/2 \max_{x\in \mathbb{X}_{\nu, \epsilon}}||x||^2$, which follows directly from the definition of the cost function (see Lemma 7.1 of \cite{Nedic2011}), then~\eqref{eq.tech1} can be written as
\begin{equation*}
||f^{*} - f^\opt||\leq M_f M_\mu \sqrt{\frac{\epsilon}{2 \nu}} + \frac{\nu}{2}D^2
\end{equation*} 
\vskip-0.5cm $\hfill$  $\Box$

\subsection{Non-symmetric Case}

In this subsection, we consider the case of non-symmetric matrices $W$. It is not difficult to see that all the previous derivations hold true with the small modification that instead of $\lambda_{\max}(\cdot)$ we will have $\sigma_{\max}(\cdot)$, meaning the largest singular value. Unfortunately, due to the term $\sigma_{\max}(\betai W - I_N)$, the condition on $\betai$ is not solvable in closed-form. Define
\begin{equation*}
C' := \max(\betai \sigma_{\max}(W), 1), \quad \kappa' := \varphi - {F_{\Phi}}  \sigma_{\max}(\beta W - I_N)
\end{equation*}
then the conditions in~\eqref{eq:conditions} for the non-symmetric case become
\begin{equation*}
\sigma_{\max}(\beta W - I_N) < \frac{\varphi}{F_{\Phi}}, \quad \textrm{ and } \quad \alphai < \frac{2 \kappa' }{C'^2 F_{\Phi}^2}.
\end{equation*}

\subsection{Weight Design}

Instead of a unique step size $\betai$, one may consider designing the whole information exchange weight matrix $W$. For simplicity we redefine the weight matrix as $\overline{W}: = \betai W$ whose pattern is fixed by the network structure (assumed here to be symmetric) but the single entries are variables to be determined. If we use $\overline{W}$ in the iterations~\eqref{eq:itx}-\eqref{eq:itmu}, the convergence conditions on $\overline{W}$ (in addition to the one on $\alpha$) can be written as
\begin{eqnarray*}
\lambda_2(\overline{W}) > 0 & \Longleftrightarrow& \overline{W} + (1/N) \textbf{1}_N\textbf{1}_N^\top \succ 0 \\
\lambda_{\max}(\overline{W}) < 1 & \Longleftrightarrow& \overline{W} - 1 \prec 0
\end{eqnarray*}
These conditions are similar to those in~\cite{Xiao2006a}. In particular, the first condition is a connectivity condition, while the second could be interpreted as diagonal dominance.
Using the fact that $(\overline{W} + (1/N) \textbf{1}_N\textbf{1}_N^\top)^{-1}\overline{W} = (I - (1/N) \textbf{1}_N\textbf{1}_N^\top)$ and therefore $ \overline{W}(\overline{W} + (1/N) \textbf{1}_N\textbf{1}_N^\top)^{-1}\overline{W} = \overline{W} $, by Schur complement these relations can be translated into the LMI~\cite{Xiao2006a}
\begin{equation}
\left[\begin{array}{cc} \overline{W} +  (1/N)\mathbf{1}_N\mathbf{1}_N^\top & \overline{W} \\
        \overline{W} &  1 \end{array}\right] \succ 0
\end{equation}
making the weight design a centralized convex problem\footnote{For a more detailed analysis on the weight selection criteria the reader is referred to \cite{Xiao2006a}.}.

\section{A Robotic Network Application}\label{sec:application}

\begin{figure*}
\psfrag{x}{\hskip-.2cm \footnotesize $\xv_{i, (1)}$}
\psfrag{y}{\footnotesize $\xv_{i, (2)}$}
\hskip-0.5cm\includegraphics[width=0.6\textwidth]{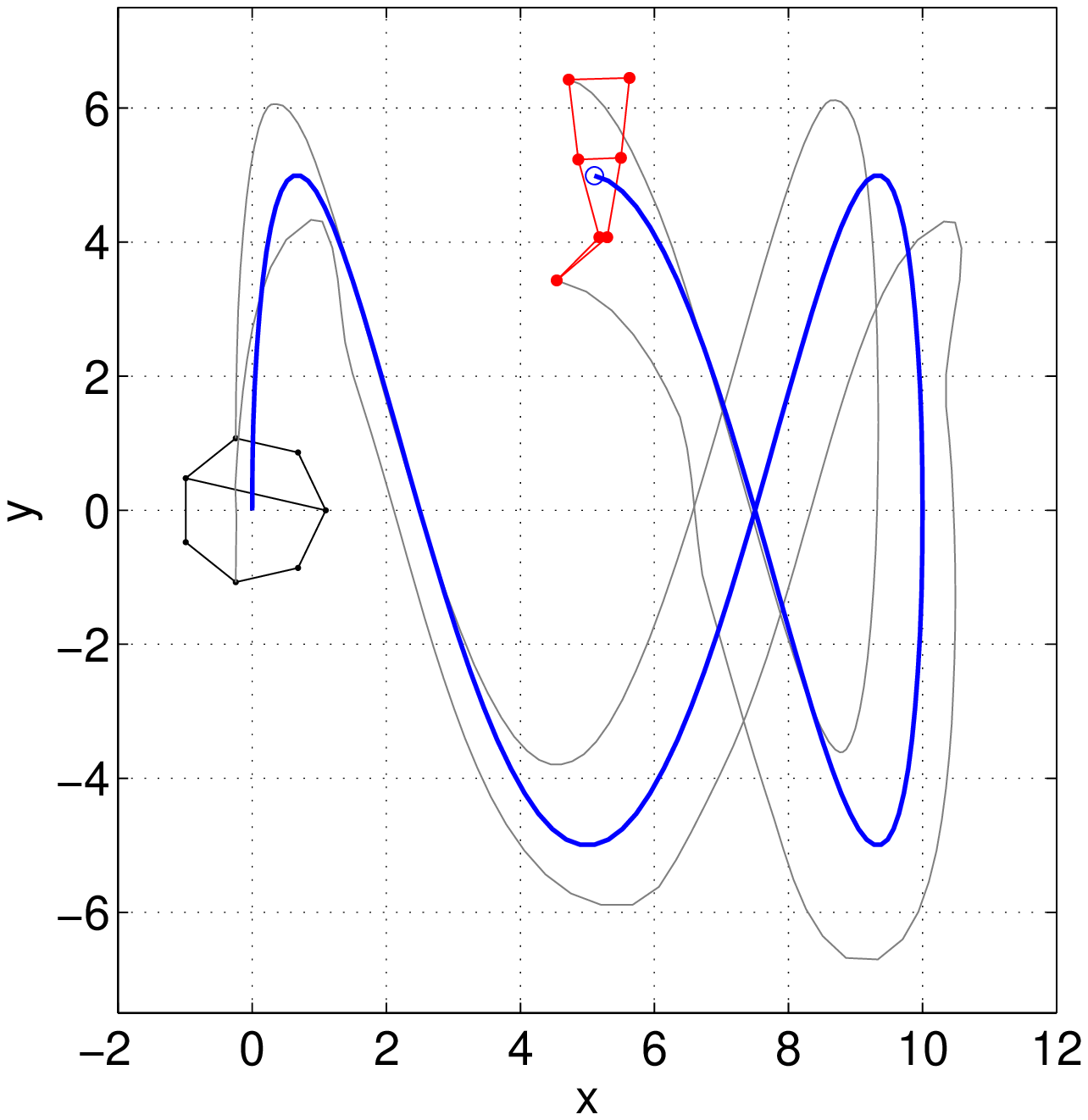}
\vskip-5.5cm{\hskip10.35cm$L \hskip-2pt= \hskip-2pt\left[\begin{array}{ccccccc} \hskip-4pt 3\hskip-4pt & \hskip-4pt -1\hskip-4pt& \hskip-4pt 0\hskip-4pt & \hskip-4pt -1\hskip-4pt&\hskip-4pt 0\hskip-4pt & \hskip-4pt 0\hskip-4pt& \hskip-4pt -1\hskip-4pt\\ %
\hskip-4pt -1\hskip-4pt & \hskip-4pt 2\hskip-4pt& \hskip-4pt -1\hskip-4pt & \hskip-4pt 0\hskip-4pt&\hskip-4pt 0\hskip-4pt & \hskip-4pt 0\hskip-4pt& \hskip-4pt 0\hskip-4pt\\%
\hskip-4pt 0\hskip-4pt & \hskip-4pt -1\hskip-4pt& \hskip-4pt 2\hskip-4pt & \hskip-4pt -1\hskip-4pt&\hskip-4pt 0\hskip-4pt & \hskip-4pt 0\hskip-4pt& \hskip-4pt 0\hskip-4pt\\
\hskip-4pt -1\hskip-4pt & \hskip-4pt 0\hskip-4pt& \hskip-4pt -1\hskip-4pt & \hskip-4pt 3\hskip-4pt&\hskip-4pt -1\hskip-4pt & \hskip-4pt 0\hskip-4pt& \hskip-4pt 0\hskip-4pt\\
\hskip-4pt 0\hskip-4pt & \hskip-4pt 0\hskip-4pt& \hskip-4pt 0\hskip-4pt & \hskip-4pt -1\hskip-4pt&\hskip-4pt 2\hskip-4pt & \hskip-4pt -1\hskip-4pt& \hskip-4pt 0\hskip-4pt\\
\hskip-4pt 0\hskip-4pt & \hskip-4pt 0\hskip-4pt& \hskip-4pt 0\hskip-4pt & \hskip-4pt 0\hskip-4pt&\hskip-4pt -1\hskip-4pt & \hskip-4pt 2\hskip-4pt& \hskip-4pt -1\hskip-4pt\\
\hskip-4pt -1\hskip-4pt & \hskip-4pt 0\hskip-4pt& \hskip-4pt 0\hskip-4pt & \hskip-4pt 0\hskip-4pt&\hskip-4pt 0\hskip-4pt & \hskip-4pt -1\hskip-4pt& \hskip-4pt 2\hskip-4pt\\
\end{array}\right]$}\vskip3cm
\caption{Representation of the trajectories of the robots while the target moves (blue thick line). The initial graph and positions of the robots are marked in black, while the final configuration is marked in red. The notation $\xv_{i, (j)}$ indicates the $j$-th component of $\xv_{i}$.}
\label{fig.traiettoria}
\end{figure*}

In this section we use an application scenario inspired by a realistic problem to illustrate the proposed method. We consider a group of $N$ mobile robots that can communicate among each other via a communication network. Let the graph that describes the network be $\mathcal{G} = (\mathcal{V}, \mathcal{E})$ and we will assume that it is time-invariant. Let $x_i(k) \in \mathbb{R}^2$ be the position of the robot $i$ at the discrete time step $k$. Let $y(k) \in \mathbb{R}^2$ be the position of a moving target at the discrete time step $k$. We model the robot dynamics as single integrator systems and associate the convex cost function $f_i(x_i(k) - x_i(k-1))$ with each of them that can represent the energy consumption. We assume the robots know the target location.

We are interested in moving the robots to ensure that the target is always in the barycenter of their positions. Furthermore, we require that robots connected by an edge in the fixed graph have a bound $R$ on their maximal distance (for communication purposes). We limit the allowable change of position in one step $||x_i(k) - x_i(k-1)||$ by $v_{\max, i}$ to model physical limitations. Finally, our global objective is to meet the aforementioned requirements while minimizing the total energy consumption. At each discrete time $k$, the above problem can be written as
\begin{eqnarray}
\minimize_{\xv_i(k), \dots, \xv_N(k)} & \sum_{i=1}^N f_i(\xv_i(k) - \xv_i(k-1)) \label{eq:robots} & \\
\subjto &||\xv_i(k) - \xv_j(k)||^2 - R^2 \leq 0 & \nonumber\\ && \hskip-1.5cm\textrm{for all } (i,j) \in \mathcal{E} \nonumber \\
&||\xv_i(k) - \xv_i(k-1)|| - v_{\max, i} \leq 0 & \nonumber \\ &&  \hskip-1.5cm\textrm{for } i = 1, \dots, N \nonumber \\        & 1/N \sum_{i = 1}^N \xv_i(k) = y(k) &\nonumber
\end{eqnarray}
which is a specific instance of~\eqref{eq:problem} for each time step $k$. Since the target is moving, $y(k)$ corresponds to a time-varying total available resources $\r$ in the formulation of problem~\eqref{eq:problem}\footnote{We note that, when the proposed saddle-point algorithm is used to solve the problem~\eqref{eq:robots} at each discrete time step $k$, each initial $\xv_i(k)^{(0)}$ can be chosen as $\xv_i(k)^{(0)} = \xv_i(k-1) + (y(k) - y(k-1))$, whereas $\xv_i(0)^{(0)} = y(0)$. This ensures that the initial iterates satisfy the resource allocation constraint. 
}.

Our simulation example consists of $N = 7$ robots connected via a communication graph shown in Figure~\ref{fig.traiettoria} with Laplacian matrix $L$. The parameters of the scenario are $R = 1.2$, and $f_i(\delta \xv_i(k)) = \langle Q_i \delta \xv_i(k), \delta \xv_i(k) \rangle$, where $\delta \xv_i(k) = \xv_i(k) - \xv_i(k-1)$ and $Q_i = 1$ for all $i$ except for $i = 6$, for which $Q_6 = 0$. We consider $v_{\max, 6} = 0.5$, while the others are set to $+\infty$. Given the fact that the cost function is not strictly convex and the position of the robots are coupled via a resource allocation constraint, even this small-size problem could be difficult to solve (in terms of communication/computation requirements) for common gradient algorithms. This makes this example interesting to analyze with the proposed approach. 

We solve problem~\eqref{eq:robots} via the regularized saddle-point algorithm with $\nu = 10$, $\epsilon = .01$, and $W = L$, $\alphai = 0.01$, and $\betai = 0.2$. Figure~\ref{fig.traiettoria} shows the computed trajectories of the robots while the target moves (blue thick line). The initial graph and positions of the robots are marked in black, while the final configuration is marked in red. In order to assess the ``quality'' of the regularized problem solution with respect to the original primal one, the maximal error of the optimal robot positions $\max_{k}||\xv^*(k) - x^\opt(k)||$ was computed and resulted in $0.02$, which is acceptable in this application scenario. Finally, we report that the total number of communication/computation iterations per discrete time step $k$ was $\kit = 2000$, and the computations required around $0.03$~s per node per discrete time step $k$, on an Intel Core \emph{i5} (2.3 GHz and 4GB DDR3) laptop. These results are encouraging since the regularization parameters were not specifically optimized to minimize the number of iterations. This aspect, along with extensive comparisons with common gradient methods are left as future development.~\footnote{As a preliminary result, we remark that dualizing the resource allocation constraint would cause an increase of the number of iterations of at least 40\%, even with $\beta = 1$. We expect non-regularized gradient methods to need even more iterations to achieve the same accuracy as comprehensively illustrated in~\cite{Nedic2011}.} 

\section{Conclusions}\label{sec:conclusions}
We have presented a regularized saddle-point algorithm that solves convex optimization problems with resource allocation constraints in a distributed fashion. Convergence conditions for the step sizes of the method were derived and their relations to the properties of the graph and the optimization problem were shown. Finally, we have illustrated our proposed algorithm with a robotic network application scenario. A more in-depth study on the optimal design of the step-sizes and the regularization parameters, more extensive simulation studies on a variety of application scenarios, as well as time-variant communication networks are among the future research directions.

%% file: appendix_v1.tex
\appendix

\subsection*{Bound on the primal iterates $\xv$} \label{section:bound}

Although we have not encountered any problems in the simulation study of Section~\ref{sec:application}, in general it is difficult to guarantee theoretically that the primal iterations $\xv^{(\kit)}$ stay bounded for all $\kit$. This is due to the fact that we cannot simply project the iterations~\eqref{eq:itx} on $\hat{\mathbb{\XV}}$ without destroying the properties of the information exchange matrix $W$. However, we show here that it is possible to find a simple way to enforce the boundedness of the iterates by the closed, convex, and bounded set $\hat{\mathbb{\XV}}$. We remark that this boundedness condition plays the important role of ensuring that the quantity based on $\xv$ stays finite. Define $\hat{\mathbb{\XV}}$ as the ball
\begin{equation*}
\hat{\mathbb{\XV}} = \displaystyle \bigcup_{i=1}^N \left\{\xv_i|\,\, ||\xv_i|| \leq \hat{\XV}_i \right\}
\end{equation*}
Consider the simple modification of the regularized Lagrangian-type function $\La(\xv, \m)$, for some scalar $\qq > 0$, as
\begin{equation}
\hat{\La}(\xv, \m, \qq) = \La(\xv, \m) - (1/\qq)\sum_{i=1}^N \log(\hat{\XV}_i-||\xv_i||)
\end{equation}
where we use a logarithmic barrier to enforce the ball constraint. Moreover, consider the modified problem:
\begin{eqnarray}
\max_\m \min_\xv & \hat{\La}(\xv,\m,\qq) \label{eq:mod}\\
\subjto & \left(\mathbf{1}_N^\top\otimes I_n\right) \xv = \r \nonumber
\end{eqnarray}
Assume that the following reasonable conditions hold:
\begin{enumerate}
\item[C1)] The first iterate $\xv^{(0)}$ is in $\hat{\mathbb{\XV}}$;
\item[C2)] The set ${\mathbb{\XV}_{\nu, \epsilon}}$ is a subset of $\hat{\mathbb{\XV}}$ and the following inequalities hold:
\begin{equation*}
\max_{\xv\in{\mathbb{\XV}_{\nu, \epsilon}}}||\xv|| \ll N \hat{\XV}_i \ll \qq
\end{equation*}
\item[C3)] the step-size $\alphai$ is \emph{sufficiently small} to guarantee that $\log(\hat{\XV}_i-||\xv^{(\kit)}_i||)$ stays finite for any $\kit$.
\end{enumerate}

\begin{theorem}
Under Conditions C1-C3 and using the modified function $\hat{\La}(\xv, \m, \qq)$ instead of $\La(\xv, \m)$ in the iterates~\eqref{eq:itx}-\eqref{eq:itmu}, the primal iterates $\xv^{(\kit)}$ are contained in the closed and bounded convex set $\hat{\mathbb{\XV}}$. In addition,
\begin{enumerate}
\item[\emph{i)}] the optimizer of the modified problem~\eqref{eq:mod} is approximately the same as the optimizer $\xv^*$ of the original problem~\eqref{eq:reg};
\item[\emph{ii)}] the difference between the optimal value of the modified problem~\eqref{eq:mod}, $\hat{\La}^*$ and the optimal value $\La^*$ of the original problem~\eqref{eq:reg} is approximately
\begin{equation}
\hat{\La}^* - {\La}^* \approx -\sum_{i=1}^N\log(\hat{\XV}_i)/ \qq
\end{equation}
which goes to zero when $t \rightarrow \infty$. 
\end{enumerate}
\end{theorem}
\emph{Proof.} Conditions C1-C3 immediately imply that the primal iterates $x^{(\kit)}$ are bounded in the closed, convex, and bounded set $\hat{\mathbb{\XV}}$. Consider now the optimal point $(\xv^*, \m^*)$ of the original problem~\eqref{eq:reg}, for which it holds that
\begin{multline*}
\hat{\La}(\xv^*, \m^*, \qq) = \La(\xv^*, \m^*) - (1/\qq)\sum_{i=1}^N \log(\hat{\XV}_i-||\xv^*_i||) \\ \displaystyle \approx_{\textrm{(C2)}}
\La(\xv^*, \m^*) - \sum_{i=1}^N\log(\hat{\XV}_i)/ \qq
\end{multline*}
from which the claims \emph{i)} and \emph{ii)} of the theorem follows. \hfill $\Box$ 